\documentclass[prb,letterpaper,aps,floatfix,twocolumn]{revtex4-1}
\usepackage{graphicx}
\usepackage{amsmath}
\usepackage{subfigure}
\newcommand{\beq}{\begin{equation}}
\newcommand{\eeq}{\end{equation}}
\newcommand{\bk}{{{\bf{k}}}}

\newcommand{\br}{{{\bf{r}}}}

\newcommand{\bB}{{\bf{B}}}
\newcommand{\bE}{{\bf{E}}}

\newcommand{\bb}{{\bf{b}}}

\newcommand{\beqa}{\begin{eqnarray}}
\newcommand{\eeqa}{\end{eqnarray}}

\newcommand{\bsigma}{{\boldsymbol \sigma}}
\newcommand{\btau}{{\boldsymbol \tau}}
\newcommand{\bOmega}{{\boldsymbol \Omega}}

\begin{document}
\title{Topological Semimetals}
\author{A.A. Burkov}
\affiliation{Department of Physics and Astronomy, University of Waterloo, Waterloo, Ontario 
N2L 3G1, Canada} 
\affiliation{ITMO University, Saint Petersburg 197101, Russia}
\date{\today}
\begin{abstract}
Topological semimetals and metals have emerged as a new frontier in the field of quantum materials. 
Novel macroscopic quantum phenomena they exhibit are not only of fundamental interest, but may hold 
some potential for technological applications.  
\end{abstract}
\maketitle
The study of the electronic structure topology of crystalline materials has emerged in the last 
decade as a major new theme in the modern condensed matter physics. 
The starting impetus came from the remarkable discovery of topological insulators,~\cite{Hasan10,Qi11} but the focus has recently 
shifted towards topological semimetals and even metals. 
While the idea that metals can have a topologically nontrivial electronic structure is not entirely new and some of the recent 
developments were anticipated in earlier work,~\cite{Volovik03,Haldane04,Murakami07} 
this shift was precipitated by the theoretical discovery of Weyl~\cite{Wan11,Ran11,Burkov11-1,Xu11,Krempa12,XiDai15,Huang15} and later Dirac semimetals.~\cite{Kane12,Fang12,Fang13}
The experimental realization of both Weyl and Dirac semimetals~\cite{Chen14,Neupane14,HasanTaAs,DingTaAs,Lu15}
within the last couple of years has brought the field to the forefront of quantum condensed matter research. 

\section{Weyl semimetals}
\label{sec:1}
While ordinary metals owe most of their unique observable character to the existence of a Fermi surface, and insulators 
to lack thereof and a finite gap between the highest occupied and the lowest unoccupied state, the defining feature of topological 
semimetals is the appearance of band touching points or nodes at the Fermi energy, where two or more bands are exactly degenerate at particular values of the crystal momentum in the first Brillouin zone (BZ). 
There may also exist line nodes, where the bands are degenerate along closed lines in momentum space. 
While the existence of such band touching nodes was recognized since the early days of solid state physics,~\cite{Herring37}
their importance was only appreciated recently. 

Naively, a band touching point should be a very unlikely and a very unstable feature, and can thus be hardly expected to be of any importance. 
Indeed, as we know from basic quantum mechanics, a degeneracy between energy levels is always lifted unless required by 
a symmetry. 
However, this naive viewpoint overlooks the possibility of an accidental degeneracy of a single pair of bands in a 
three dimensional material. 
To see how this happens, suppose two bands touch at some point $\bk_0$ in the first BZ and at energy $\epsilon_0$.
In the vicinity of this point, the momentum-space Hamiltonian may be expanded in Taylor series with respect to the  
deviation of the crystal momentum from $\bk_0$. 
The expansion will generally have the following form (ignoring possible anisotropies for simplicity)
\beq
\label{eq:1}
H(\bk) = \epsilon_0 \sigma_0 \pm \hbar v_F (\bk - \bk_0) \cdot \bsigma, 
\eeq 
where $\sigma_0$ is a $2 \times 2$ unit matrix and $\bsigma = (\sigma^x, \sigma^y, \sigma^z)$ are the three Pauli matrices. 
This represents nothing more than an expansion of a general $2\times 2$ Hermitian matrix in terms of the unit matrix and the three Pauli matrices.  
What makes this interesting is that there is nothing we can do to Eq.~\eqref{eq:1} to get rid of the band touching point.
Changing $\epsilon_0$ or $\bk_0$ can only change the location of the band touching point in energy or crystal momentum, 
while changing the parameter $v_F$, which has dimensions of velocity, only changes the slope of the band dispersion away from 
the point. The point itself is always there. 
Formally, this has to do with the fact that the number of crystal momentum components is the same as the number of the Pauli 
matrices, both being three. 
This means that for such an unremovable band touching to occur we need three spatial dimensions, which is not a 
problem, and nondegenerate (at a general value of the crystal momentum) bands. 
This second requirement can not be satisfied in a material that possesses two fundamental symmetries: inversion $P$ (which means that the crystal structure has an inversion center) and time reversal $\Theta$ (the material is nonmagnetic). 
The reason is that in such a material all bands must be at least doubly degenerate at every value of $\bk$ due to the 
fundamental property of any system of fermions that $(P \Theta)^2 = -1$. 
Thus we come to the conclusion that unremovable band touching points may only occur in noncentrosymmetric or magnetic materials.

Interestingly, if we set $\epsilon_0 = 0$ in Eq.~\eqref{eq:1}, which simply resets the zero of 
energy, Eq.~\eqref{eq:1} takes the exact form (up to a trivial replacement of the speed of light by $v_F$) of a Weyl Hamiltonian, 
that is the Hamiltonian of a massless relativistic particle of right-handed or left-handed chirality, which corresponds to the $\pm$ 
sign in front. 
Weyl fermions are fundamental building blocks, from which the Standard Model of particle physics is constructed. 
All the known elementary fermions in nature have masses, acquired due to complex interactions between the fundamental 
Weyl fermions. The fact that massless Weyl fermions may occur as quasiparticles in a crystal is very intriguing and 
may have deep implications for our understanding of the Universe.~\cite{Volovik03}
Due to this analogy with the Weyl equation, the touching points of pairs of nondegenerate bands are called Weyl points 
or nodes. 

Another important feature of the Weyl Hamiltonian is that it is a topologically nontrivial object. 
Indeed, the eigenstates of Eq.~\eqref{eq:1} may be labelled by {\em helicity}, which is the sign of the projection of  
$\bsigma$ onto the direction of the crystal momentum $\bk$ (helicity coincides with chirality for a massless 
particle, but not in general). 
The expectation value of $\bsigma$ in an eigenstate of a given helicity forms a vector field in momentum space that 
wraps around the Weyl node location $\bk_0$, forming a ``hedgehog", or a ``hairy ball". 
Such a ``hedgehog" may be characterized by a topological invariant, defined as a flux of a vector 
\beq
\label{eq:2}
\bOmega(\bk) = \pm \frac{\bk}{2 |\bk|^3}, 
\eeq
through any surface in momentum space, enclosing the Weyl node, where the sign in front is the chirality of the node. 
This flux, normalized by $2\pi$, is equal to the chirality, and therefore this is a quantized integer-valued invariant. 
The vector $\bOmega(\bk)$ is called Berry curvature and the Weyl nodes may thus be regarded as point ``charges", which 
are sources and sinks of the Berry curvature. 
This implies that the Weyl nodes must occur in pairs of opposite chirality, since 
the field lines of the Berry curvature must begin and end somewhere within the BZ, and that the only way to eliminate the Weyl nodes is to annihilate them pairwise, which may only happen if they occur at the same point in momentum space.
Thus the origin of the stability of the Weyl nodes is ultimately topological. 
\begin{figure}[t]
\vspace{-2cm}
\includegraphics[width=8cm]{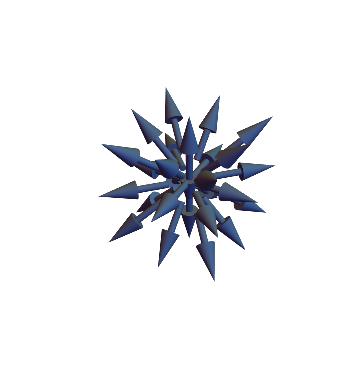}
\vspace{-2cm}
\caption{Berry curvature field forming a ``hedgehog" around a Weyl node.}  
  \label{fig:hedgehog}
\end{figure} 

An important consequence of the fact that Weyl nodes are point sources and sinks of the Berry curvature field is that 
when the sample has edges in real space, parallel to the line, connecting a pair of Weyl nodes in momentum 
space, there exist edge-localized states, which are confined to the projection of the internode interval onto the first 
BZ of the sample surface. These edge states are called Fermi arcs.~\cite{Wan11}

So far we have deduced that any three-dimensional noncentrosymmetric or magnetic material will have Weyl 
band touching points somewhere in its electronic structure. 
Yet in order for these points to have a significant effect on observable properties of the material they must 
occur at or very close to the Fermi energy and there must not be any other states at the Fermi energy. 
This is what is nontrivial to achieve in general. 
What may help here is the fact that any nondegenerate band must have exactly the same number of states, equal 
to the number of unit cells in the crystal and the Pauli principle, which prohibits double occupation of any state. 
What this implies is that, for any even integer number of electrons per unit cell
the bands are always either completely filled or completely empty, 
provided they do not overlap in energy. They problem is that they generally do overlap, but we may find a situation 
when they do not if we look among materials which are called narrow gap semiconductors, which in the absence of 
broken inversion or time reversal symmetry would be insulators with a very small gap. 
All currently known Weyl materials are of this sort.   
\begin{figure}[t]
\subfigure{
  \includegraphics[width=8cm]{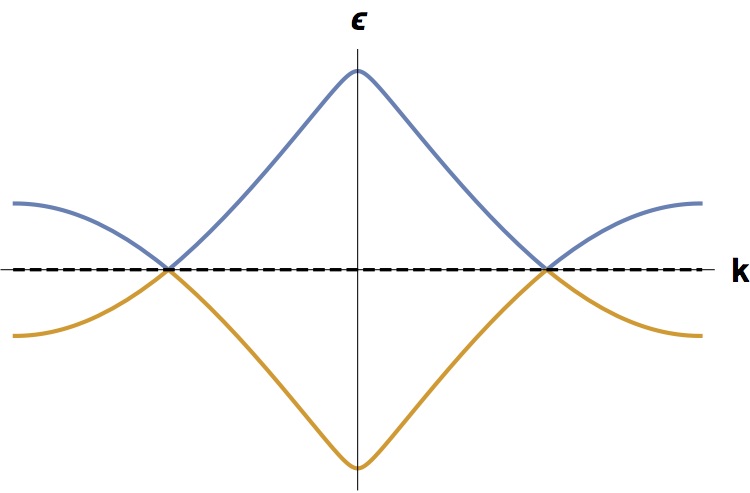}}
  \subfigure{
  \includegraphics[width=8.5cm]{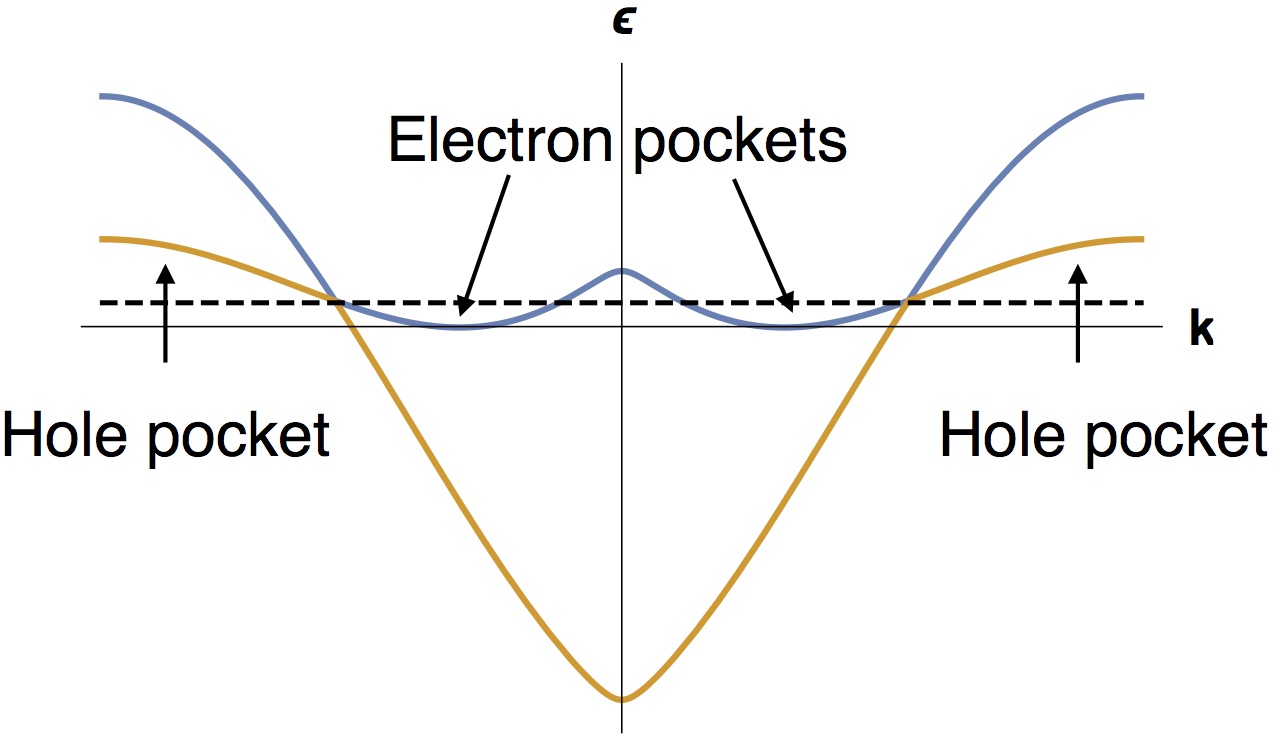}}
  \caption{Type-I (top panel) versus type-II (bottom panel) Weyl nodes. In the type-I case, the Fermi energy, represented by the dashed line, only crosses the nodes themselves and is otherwise in the gap between the two touching bands. In the type-II case there are electron and hole pockets, that touch at the Weyl nodes.}  
  \label{fig:1}
\end{figure} 

Eq.~\eqref{eq:1}, as written, actually makes an implicit omission, which leads to a loss of important piece of physics 
of Weyl semimetals. In general, the coefficient of the unit matrix $\sigma^0$ is a function of the crystal momentum, 
not just a constant $\epsilon_0$, which is simply the zeroth order term in the Taylor expansion of the function.
An important question is whether this expansion has a term, linear in $\bk - \bk_0$, or not. 
Only if $\bk_0$ is a special time reversal invariant momentum (TRIM), which occur only at the 
BZ center or at special points at the BZ boundary, there is no linear term. In general, Weyl points occur away from TRIM, which means that a linear term does exist
\beq
\label{eq:3}
H = \hbar \tilde v_F (\bk - \bk_0) \sigma_0 \pm \hbar v_F (\bk - \bk_0) \cdot \bsigma, 
\eeq
where we have set $\epsilon_0 = 0$.
Now, something new happens when $\tilde v_F > v_F$. The Weyl point is of course still there, but the two bands that 
touch at the Weyl point now overlap in energy, forming electron and hole pockets.
The Weyl point then becomes a point at which an electron and a hole pocket touch. 
Such Weyl semimetals are called type-II Weyl semimetals,~\cite{Soluyanov15} see Fig.~\ref{fig:1}. 
MoTe$_2$ is an example of a type-II Weyl semimetal material.~\cite{Kaminski16}
TaAs, the first discovered Weyl semimetal material,~\cite{HasanTaAs,DingTaAs} is type-I. 

An additional subclass of Weyl semimetal materials are the so-called multi-Weyl semimetals,~\cite{multiweyl}
in which Weyl nodes carry topological charges of higher magnitude (e.g. 2 or 3). Such higher-charge Weyl nodes 
may be stabilized by certain point-group crystal symmetries. 
\section{Dirac semimetals}
\label{sec:2}
As mentioned above, we may imagine getting a Weyl semimetal by breaking inversion or time reversal 
symmetry in a narrow gap semiconductor. 
The simplest Hamiltonian that describes this situation has the form
\beqa
\label{eq:4}
H = \left(
\begin{array}{cc}
\hbar v_F \bsigma \cdot \bk & m \\
m & - \hbar v_F \bsigma \cdot \bk
\end{array}
\right). 
\eeqa
This describes two Weyl fermions of opposite chirality at the same point (TRIM) in the BZ. 
The off-diagonal entry $m$ in Eq.~\eqref{eq:4} mixes the two Weyl fermions and opens up a gap of magnitude $2m$. 
This Hamiltonian in fact describes a transition between a topological and ordinary insulator in three dimensions. 
The transition point corresponds to setting $m = 0$ and is what is called a Dirac semimetal. 
At the Dirac point we have degeneracy between four bands instead of two at a Weyl point. Such a degeneracy 
requires either fine tuning or an extra symmetry to force $m = 0$. 
In the case when $m = 0$ by symmetry we get a stable Dirac semimetal phase, distinct from Weyl semimetal. 
It can be shown that symmetry-protected Dirac points of the kind described above, may only occur at TRIM at the BZ edge 
and require a nonsymmorphic symmetry, which means a rotation combined with a partial translation.~\cite{Kane12}
While proposals for specific material realizations have been made, no such Dirac semimetal has yet been realized experimentally.

However, another possibility exists.  Instead of a single Dirac point at a TRIM, it is possible to create two Dirac points 
on a rotation axis, protected by a rotational symmetry about this axis. 
The Hamiltonian that describes this situation has the form 
\beq
\label{eq:5}
H = \hbar v_F(\tau^x \sigma^z k_x - \tau^y k_y) + m(k_z) \tau^z, 
\eeq
where $\btau$ and $\bsigma$ are Pauli matrices, describing orbital and spin degrees of freedom correspondingly 
and $m(k_z) = -m_0 + m_1 k_z ^2$. 
For a given eigenvalue of the spin operator $\sigma = \pm 1$, this Hamiltonian describes the simplest Weyl semimetal with 
two Weyl nodes at points $\bk_{\pm} = (0, 0, \pm \sqrt{m_0/ m_1})$. 
The two Weyl semimetals, corresponding to $\sigma = \pm 1$, are related to each other by the time reversal operation. 
Thus each of the two band touching points $\bk_{\pm}$ contain two Weyl points of opposite chirality and they are 
therefore two Dirac points. 
This kind of Dirac semimetal is distinct from the one, discussed above, with a single Dirac point at a TRIM. 
Such a Dirac semimetal has been realized experimentally in two compounds, Na$_3$Bi and Cd$_2$As$_3$.~\cite{Fang12,Fang13,Chen14,Neupane14}
\section{Topological response}
\label{sec:3}
While the existence of special edge states, required by the electronic structure topology, is a feature of Weyl and 
Dirac semimetals that is similar to the other class of topological materials, the topological insulators, what makes
the former potentially more interesting is that they exhibit special topology-related response, in addition to special 
spectroscopic features like the edge states. 

The best known example of a universal (namely detail-independent) topological transport property is the Hall conductivity of a two-dimensional 
quantum Hall liquid, which is quantized to an incredible precision, in spite of all the 
imperfections and dirt in the semiconductor heterostructure samples, in which it is measured. 
What leads to this precision and universality is that a two-dimensional quantum Hall liquid is a (topological) 
insulator and it is the intrinsic insensitivity of an insulator to perturbations that is the source of it. 
One would not expect such a universal transport to exist in a gapless system. 

Surprisingly, Weyl and Dirac semimetals do exhibit universal transport. 
Its origin is the chiral anomaly, a phenomenon that was discovered many years ago in the particle physics 
context. 
In this case chiral anomaly is an unexpected nonconservation of chiral charge 
in a system of massless relativistic fermions, coupled to electromagnetic field with collinear electric and 
magnetic components. 
It is unexpected because being massless for a relativistic particle, described by the Dirac equation, is equivalent to having a conserved chirality, or chiral symmetry. 
The Dirac equation also has negative energy state solutions, which must be filled with unobservable particles, forming the 
so-called Dirac sea. 
The seemingly obvious chiral symmetry disappears, however, once the infinite negative energy Dirac sea is filled with 
particles, and it is the infinity of the Dirac sea, which by itself is surely a mathematical artifact of the relativistic field 
theory, that ultimately leads to the anomaly. 
Astonishingly, the  chiral anomaly is nevertheless a real effect, with observable consequences, explaining for example the decay
of a neutral pion into two photons.
Mathematically, chiral anomaly may be expressed most conveniently as the following extra term in the action for the 
electromagnetic field, induced by the fermions
\beqa
\label{eq:6}
S&=&\frac{e^2}{32 \pi^2 \hbar^2 c} \int d t d^3 r \theta(\br, t) \epsilon^{\mu \nu \alpha \beta} F_{\mu \nu} F_{\alpha \beta}
\nonumber \\
&=&\frac{e^2}{8 \pi^2 \hbar^2 c} \int d t d^3 r \theta(\br, t) \bE \cdot \bB,
\eeqa
where $\theta(\br, t) = 2 (\bb \cdot \br - b_0 t)$ and $b^\mu = (b_0,\bb)$ are chiral gauge fields, which couple antisymmetrically to 
Weyl fermions of different chirality, while $F_{\mu \nu} = \partial_\mu A_\nu - \partial_\nu A_\mu$ is the electromagnetic 
field tensor and $A_\mu$ are the ordinary electromagnetic gauge fields, which do not care about chirality. 
\begin{figure}[t]
  \includegraphics[width=8cm]{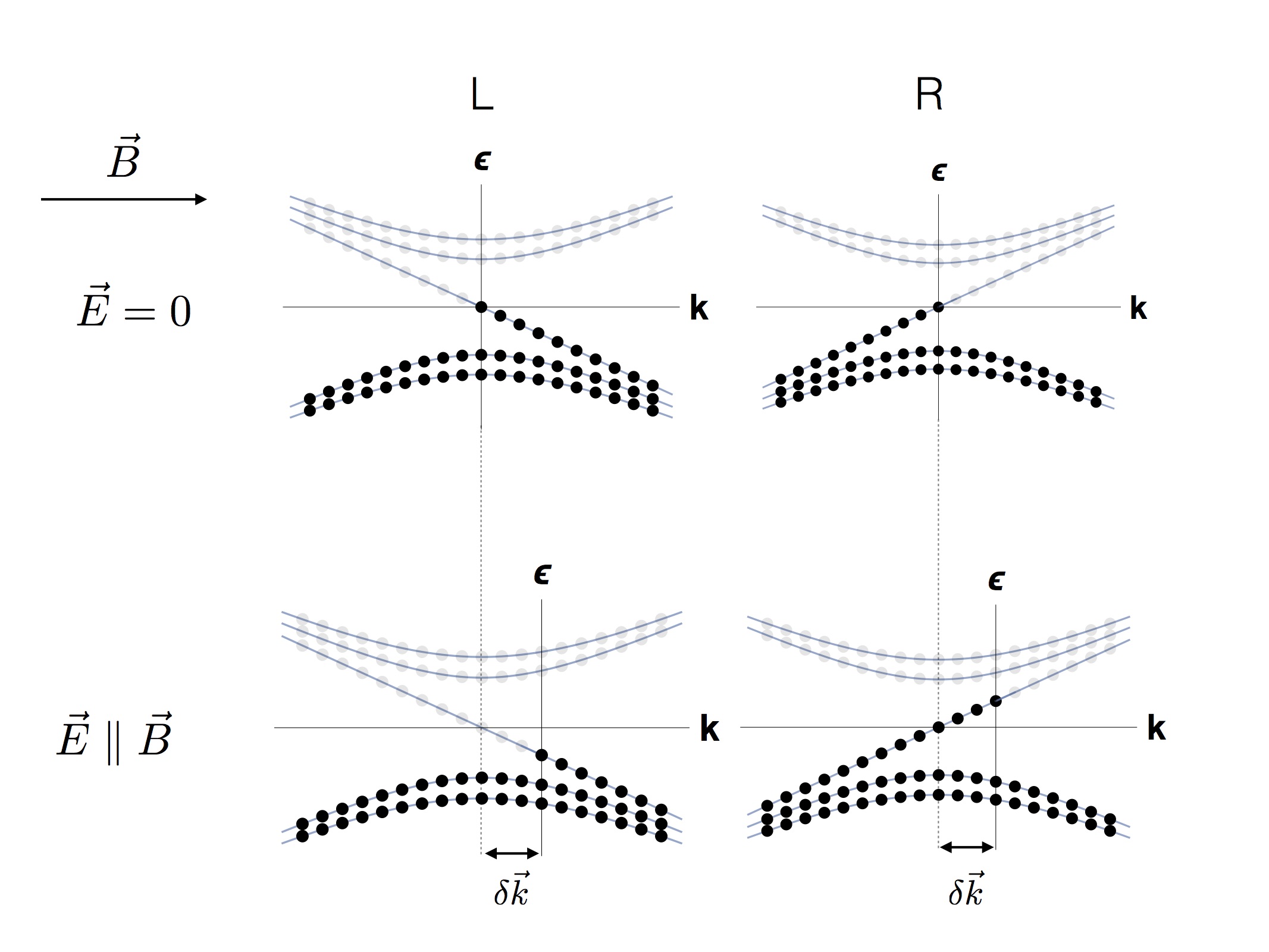}
  \caption{Simple explanation of the chiral anomaly, based on Landau level spectrum of Weyl fermions in an external magnetic field. Top panel: Energy spectrum of the left-handed (L) and the right-handed (R) fermions in equilibrium in the presence of a magnetic field $\vec B$. Filled states with negative energy are shown by black dots, while empty states with positive energy by gray dots. Bottom panel: Same spectrum, but in the presence, in addition, of an electric field $\vec E$, parallel to the magnetic field $\vec B$. 
 All states have been displaced in momentum space by an amount $\delta \vec k \sim - \vec E$ from their equilibrium locations. 
 As a consequence, right-handed particles and left-handed antiparticles have been produced.}
  \label{fig:2}
\end{figure} 
In the Weyl and Dirac semimetal context what leads to the analogous phenomena is the emergence of an effective 
relativistic structure in the Hamiltonian near Weyl or Dirac band touching points, as described above. 
In this case, chiral symmetry is really not there to begin with and there is thus nothing ``anomalous" in the 
chiral charge nonconservation. 
The astonishment is still there, however, but the reason for it is opposite: in spite of chirality not being strictly conserved, 
the anomaly term looks exactly like Eq.~\eqref{eq:6}, as long as the Fermi energy is close to the nodes!
In this context
\beq
\label{eq:7}
\bb = \frac{1}{2} \sum_i C_i \bk_i, \,\, b_0 = \frac{1}{2} \sum_i C_i \epsilon_i, 
\eeq
where $C_i$ is the topological charge (chirality) of the Weyl node $i$, $\bk_i$ is its location in momentum space 
and $\epsilon_i$ is its energy. 
This leads to universal transport phenomena, such as anomalous Hall effect, whose magnitude is determined only by the 
location of the Weyl nodes in magnetic Weyl semimetals; and a large negative longitudinal magnetoresistance in both Weyl 
and Dirac semimetals.~\cite{Spivak12,Burkov_lmr_prb,Ong_anomaly,Burkov_Z2} The latter effect arises as a result of a two-step 
process: generation of a chiral chemical potential by charge current in the presence of an external magnetic field, which in turn  
leads to an extra contribution to the charge current, known as the Chiral Magnetic Effect.~\cite{Kharzeev08,Franz13,Chen13}
As both steps involve the magnetic field, the resulting magnetoresistance depends quadratically on it. 
Chiral anomaly in Weyl and Dirac semimetals is a perfect example of an emergent phenomenon, characterized by simplicity and universality emerging at low energies, in this particular example due to nontrivial electronic structure topology. 

While we limited our discussion above to the electrical transport phenomena, chiral anomaly in fact affects 
thermoelectric response as well.~\cite{GdPtBi,Fiete16,Spivak16}
Many of the topological insulator and topological semimetal materials exhibit large thermopower, which by itself 
is of significant interest from the viewpoint of practical applications. 
Chiral anomaly, in addition, leads to a strong magnetic field dependence of the thermoelectric transport coefficients. 
\section{Outlook}
\label{sec:4}
Perhaps the most important current issue in the field of topological semimetals is finding more and better 
material realizations. 
As described in the previous section, universality of the topological transport properties of Weyl and Dirac semimetals is 
an emergent property, which becomes more and more precise as the Fermi energy is approaching the band touching 
points. 
Thus it is important to search for materials with the Fermi energy very close to the nodes and no other states at it. 
Of the currently known materials, Na$_3$Bi comes closest to this ideal, and it is not an accident that it exhibits 
a very strong negative longitudinal magnetoresistance due to the chiral anomaly.~\cite{Ong_anomaly,Burkov_lmr_prb,Burkov_Z2}

Another important direction is to look for a magnetic Weyl semimetal material, which would exhibit another consequence
of the chiral anomaly, the universal anomalous Hall effect. 
A material exhibiting a quantum anomalous Hall effect at room temperature would be of great technological importance, 
offering low-dissipation transport in the absence of external magnetic fields. 
This could hold some potential for helping us to avoid the quickly approaching technological brick wall: the current semiconductor electronics is nearly at its limits for both miniaturization and speed. 
Unless a fundamentally new technology emerges, our computers will very soon, within a few years,  
stop getting smaller and faster, something that we have learned to take for granted, but which in reality is not guaranteed at all. 
Materials with nontrivial electronic structure topology, exhibiting low-dissipation transport at room temperature, may help us overcome this challenge by dramatically reducing power consumption and heating. 

Finally, it is of interest, both from the fundamental and the practical standpoint, to explore the possible effects of electron-electron 
interactions in topological semimetals. 
At the simplest (in theory) level, superconductivity, resulting from attractive electron-electron interactions, will arise at low temperatures in almost any metal. 
In Weyl metals, especially in magnetic ones, superconductivity is predicted to be topologically-nontrivial, leading to Majorana 
surface states,~\cite{Bednik15,YiLi15} which may be of interest for topological quantum computing. 
Venturing into a more exotic area, one could speculate if sufficiently strong repulsive interactions could lead to fractionalized Weyl and Dirac 
semimetals, in analogy to fractional topological insulators.~\cite{FTI,Grushin16}
\begin{acknowledgments}
Financial support was provided by NSERC of Canada. 
\end{acknowledgments}
\bibliography{references}
\end{document}